\documentclass[ aps, prd, twocolumn, superscriptaddress, preprintnumbers, nofootinbib, longbibliography]{revtex4-1}

\usepackage{amsmath}	
\usepackage{amsthm}		
\usepackage{amssymb}	
\usepackage{datetime}
\usepackage{graphicx}
\usepackage{color}
\usepackage{verbatim}
\usepackage{bm}

\usepackage[normalem]{ulem}

\usepackage[notrig]{physics}

\usepackage[colorlinks=true, citecolor=midblue, linkcolor=midblue, urlcolor=midblue]{hyperref}

\usepackage{setspace}

\usepackage[svgnames]{xcolor}

\usepackage{tikz,xcolor}
\definecolor{lime}{HTML}{A6CE39}
\DeclareRobustCommand{\orcidicon}{
	\begin{tikzpicture}
	\draw[lime, fill=lime] (0,0) 
	circle [radius=0.16] 
	node[white] {{\fontfamily{qag}\selectfont \tiny ID}};
	\draw[white, fill=white] (-0.0625,0.095) 
	circle [radius=0.007];
	\end{tikzpicture}
	\hspace{-2mm}
}
\foreach \x in {A, ..., Z}{
	\expandafter\xdef\csname orcid\x\endcsname{\noexpand\href{https://orcid.org/\csname orcidauthor\x\endcsname}{\noexpand\orcidicon}}
}

\definecolor{grey}{rgb}{0.4,0.4,0.4}
\definecolor{dullmagenta}{rgb}{0.4,0,0.4}
\definecolor{darkblue}{rgb}{0,0,0.4}
\definecolor{midblue}{rgb}{0,0,0.5}
\definecolor{midred}{rgb}{0.5,0,0}
\definecolor{orange}{rgb}{1,0.5,0}
\definecolor{lightbrown}{rgb}{0.75,0.5,0.25}
\definecolor{tan}{cmyk}{0.14,0.42,0.56,0}
\definecolor{djunglegreen}{cmyk}{0.99,0,0.52,0}
\definecolor{lightgreen}{rgb}{0,1,0}
\definecolor{olivegreen}{cmyk}{0.64,0,0.95,0.40}
\definecolor{midgreen}{rgb}{0.0,0.675,0.0}
\definecolor{darkgreen}{rgb}{0,0.5,0}

\settimeformat{ampmtime}

\settimeformat{ampmtime}

\usepackage[font=small,labelfont=bf,justification=raggedright]{caption}
\usepackage{subcaption}

\newcommand{\FirstAffiliation}{\affiliation{
	Arnold Sommerfeld Center,
	Ludwig-Maximilians-Universit{\"a}t,
	Theresienstra{\ss}e 37,
	80333 M{\"u}nchen,
	Germany}}
 
\newcommand{\SecondAffiliation}{\affiliation{
	Max-Planck-Institut f{\"u}r Physik,
	Boltzmannstra{\ss}e 8,
	85748 Garching,
	Germany}}

\newcommand{\FourthAffiliation}{\affiliation{
Institut de F\'isica d’Altes Energies (IFAE) and The Barcelona Institute of Science and Technology (BIST),
Campus UAB, 08193 Bellaterra (Barcelona), Spain}}

\date{\formatdate{\day}{\month}{\year}, \currenttime}

\begin{document}

\title{
Witten Effect in $3$-Form Description of
$\theta$-vacua 
}

\author{Maximilian Bachmaier\orcidB{}}
\email{maximilian.bachmaier@physik.uni-muenchen.de}
\FirstAffiliation
\SecondAffiliation

\author{Gia Dvali}
\FirstAffiliation
\SecondAffiliation

\author{Juan Sebasti\'an Valbuena-Berm\'udez\orcidA{}}
\email{jvalbuena@ifae.es}
\FourthAffiliation

\date{\small\today} 

\begin{abstract} 

\noindent 

The $\theta$-vacua of a gauge theory admit an equivalent formulation as vacua of a massless Chern-Simons $3$-form,  which originate from the topological susceptibility of the vacuum. 
 This formulation provides a framework in which the physical manifestations of the $\theta$-angle, which are quantum in origin, can be captured at the level of effective classical equations of motion. Within this framework, we derive the Witten effect, demonstrating that in the background of a massless $3$-form, the magnetic monopole indeed acquires an electric charge proportional to $\theta$. This result, in particular, provides evidence that instantons, even when constrained by the Higgs effect, maintain a non-zero topological susceptibility of the vacuum.
In addition to the Witten effect, we numerically demonstrate that a magnetic monopole exhibits polarizability when placed in a constant background electric field.

\end{abstract}

\maketitle
    
\section{Introduction}
\label{sec:introduction}

Within a medium characterized by a nonzero $\theta$-angle, monopoles acquire an electric charge proportional to this angle. 
Witten analyzed this effect in an $SU(2)$ gauge theory with an adjoint matter field $\phi$ that spontaneously breaks (``Higgses") the gauge symmetry to $U(1)$~\cite{Witten:1979ey}. There, the Noether charge operator $N$, associated with the unbroken $U(1)$ symmetry
consists of the electric charge and an additional term proportional to the $\theta$-angle and the magnetic charge: 
\begin{align}
\label{eq:Witten-charge}
    N=\frac{1}{g}Q_e-\frac{\theta g}{8\pi^2}Q_m,
\end{align}
where for the magnetic monopole $Q_m=4\pi/g$, with $g$ being the coupling constant.
Quantization of this Noether charge leads to the modification of the Dirac-Schwinger-Zwanziger condition for the electric charge
\begin{align}
    q_e=n g+\frac{\theta g}{2\pi},
\end{align}
where $n$ is an integer.
Therefore, a magnetic monopole in a $\theta$-vacuum always carries a Witten electric charge $Q_W=\theta g /2\pi.$

Notice that the Witten effect also appears classically within the framework of axion electrodynamics, where the electric charge is induced by the gradient of the axion profile~\cite{Wilczek:1987mv}. The monopole's electric charge can also emerge dynamically, for instance, when the monopole passes through an axionic domain wall~\cite{Sikivie:1984yz}.

In the present paper, we shall adopt 
the following approach to the Witten effect. 
First, the Witten effect represents a manifestation of physically observable effects of $\theta$, which in a naive classical treatment would be disregarded as an irrelevant boundary term. However, in quantum theory $\theta$ is physical due to instantons
\cite{Callan:1976je, Jackiw:1976pf}. 

We wish to formulate a description that allows capturing 
their influence on the magnetic monopole at the level of effective classical equations of the 't Hooft-Polyakov theory~\cite{tHooft:1974kcl,Polyakov:1974ek}. 
In this, we shall rely on a 
$3$-form description of the $\theta$-vacua~\cite{Dvali:2005an} originating from the topological susceptibility of the vacuum (TSV).

It is well-known that physicality of the boundary term  $\theta G\tilde{G}$ is directly linked with the existence of the TSV correlator defined as\footnote{Notice that throughout this work, we will use $G\tilde{G}$ as a notation for $\Tr (G_{\mu\nu}\tilde{G}^{\mu\nu})$.}, 
 \begin{equation}
    \label{eq:TSV-correlator}
    \mathrm{FT} \langle G\tilde{G}(x), G\tilde{G}(0) \rangle_{p\rightarrow 0}
    = \text{const.} \neq  0 \,,
\end{equation}
Here, $\mathrm{FT}$ stands for the Fourier-transform, 
and $p$ is the four-momentum. 
This long-range correlator mediates the physical effects of the boundary term.  
  
In~\cite{Dvali:2005an}, it was shown that the TSV allows for a description of the $\theta$-vacua in terms of the vacua of a massless $3$-form field residing in the Chern-Simons $3$-form of the gauge theory. This formulation provides a framework that allows capturing the physical effects of $\theta$ in terms of effective classical equations of motion. 
  
In the present paper, we apply this method to the Witten effect and show that in a $3$-form vacuum, a magnetic monopole indeed acquires an electric charge proportional to $\theta$. 
In particular, we insert a 't Hooft-Polyakov monopole on top of a constant $G\tilde{G}$ background and, through numerical relaxation, we find that the monopole obtains electric charge.
 
On top of verifying that the $3$-form formulation correctly captures the Witten effect, this result sheds useful light on some profound questions in instanton physics. The point is that in a theory with a magnetic monopole, the gauge $SU(2)$ symmetry is Higgsed. Due to the Higgs effect, the instantons become constrained~\cite{Affleck:1980mp, Nielsen:1999vq}.

Despite this, instantons are believed to continue generating the non-zero TSV, which keeps $\theta$ physical. A discussion of this question within the electroweak theory can be found in~\cite{Anselm:1993uj}, whereas the $3$-form formulation of the electroweak $\theta$-vacuum has been given in~\cite{Dvali:2024zpc, Dvali:2025pcx}.

Due to the connection between the TSV and the physical effects of $\theta$, a recovery of the Witten effect from the $3$-form formalism serves as independent evidence for the TSV contribution from the constrained instantons.
 
\section{$3$-Form Formulation of the $\theta$-vacua}
\label{sec:3-form-formulation-of-the-theta-vacuum}

Before implementing the Witten effect, we shall briefly review the description of the $\theta$-vacua of an $SU(N)$ gauge theory in the language of a massless $3$-form field~[\citenum{Dvali:2005an, Dvali:2005zk,
Dvali:2005ws, Dvali:2013cpa, Dvali:2017mpy, Dvali:2022fdv}, \citenum{Dvali:2024zpc, Dvali:2025pcx}]. 

In order to explain the main idea, let us first consider the simplest version of the theory of a free massless $3$-form field $C_{\nu\alpha\beta}$, 
\begin{align} \label{3free}
    \mathcal{L}=\frac{1}{2}E^2,
\end{align}
where $E=\frac{1}{3!}\varepsilon^{\mu\nu\alpha\beta}\partial_\mu C_{\nu\alpha\beta}$ is the field strength ($\varepsilon_{0123}=-1$). 
The theory exhibits the gauge redundancy under the shift 
of the $3$-form by an exterior derivative of an arbitrary 
$2$-form $\Omega_{\alpha\beta}$:
\begin{equation}
C_{\nu\alpha\beta} \rightarrow C_{\nu\alpha\beta} + 
\partial_{[\mu}\Omega_{\alpha\beta]}\, . 
\end{equation}
Due to this gauge redundancy, the massless $3$-form does not contain any propagating degree of freedom. Nevertheless, it can provide very important long-range correlations via its ``electric" field. Indeed, the equation of motion following from (\ref{3free}), 
    \begin{align}
\label{eq:KB}
    \partial_\mu E \, = \, 0 \,,
\end{align}
is solved by an arbitrary constant $E\, = \, \theta \Lambda^2$, which, for future convenience, we have parameterized as the product of a dimensionless parameter $\theta$ and the energy scale $\Lambda$. 

Notice that in the above theory, these constant values of the field define different vacuum states. Although the energies of these states are different, due to the absence of any mobile charges with respect to $C$, the value of the field $E$ cannot be changed and thereby defines a super-selection sector.

We can expand the theory about the given vacuum by applying the field redefinition $E\mapsto E+\theta \Lambda^2$. 
The $\theta$-term then appears in the Lagrangian
\begin{align}
    \mathcal{L}=\frac{1}{2}E^2 + \theta \Lambda^2 E \,,
\end{align}
in the form of a total derivative, which can be 
rewritten as a boundary term,
\begin{equation} \label{Boundary1}
   \int_{3+1} \dd^4x \, \theta \Lambda^2 E \, = \, 
   \frac{\theta \Lambda^2}{3!} \int_{2+1} \dd X^{\mu} \wedge \dd X^{\nu} \wedge \dd X^{\alpha} \, 
   C_{\mu\nu\alpha} \,,
\end{equation}
where in the last expression the integration is taken over the $2+1$-dimensional world-volume of the boundary with the embedding coordinates $X^{\mu}$.

Now, the point of~\cite{Dvali:2005an} is that 
the $\theta$-vacua of the ordinary $SU(N)$ gauge theory can be described as the above-discussed $3$-form vacua in which the role of $C$ is played by the Chern-Simons $3$-form:
\begin{equation} \label{3CC}
C_{\mu\nu\alpha}^{(W)} = \frac{2}{\Lambda^2}\Tr \left(W_{[\mu}\partial_\nu W_{\alpha]}-\frac{2}{3}i g W_{[\mu}W_\nu W_{\alpha]}\right)\,.
\end{equation}
It is clear that $G\tilde{G}$ of the $SU(N)$ theory represents an invariant ``electric" field strength $E^{(W)}$ with respect to the Chern-Simons $3$-form according 
to the definition below the equation~\eqref{3free}:
\begin{align}
    \label{eq:GGdual-E-relation}
    \Tr (G_{\mu\nu}\tilde{G}^{\mu\nu})=\Lambda^2 E^{(W)},
\end{align}
with the $SU(N)$ field strength $G_{\mu\nu}=\partial_\mu W_\nu -\partial_\nu W_\mu -ig [W_\mu, W_\nu]$ and $\tilde{G}^{\mu\nu}=\frac{1}{2}\varepsilon^{\mu\nu\alpha\beta}G_{\alpha\beta}$.
At the level of the above identification, 
$C^{(W)}$ and $E^{(W)}$ are composite operators, 
which includes a massless $3$-form field. This can be understood from the fact that the physicality of $\theta$ implies a non-vanishing TSV correlator 
\begin{equation}
    \label{eq:TSV-correlator2}
    \mathrm{FT} \langle E(x)^{(W)}, E(0)^{(W)}\rangle_{p\rightarrow 0}
    = \text{const.} \neq  0.
\end{equation}
This correlator is non-zero due to instantons. From here, it is clear that the $3$-form $C^{(W)}$ correlator contains a massless pole
\begin{equation} \label{CCcorr}
    \mathrm{FT} \langle C^{(W)}(x),C^{W}(0)\rangle_{p\rightarrow 0}
    \propto  {\frac{1}{p^2}} \, .
\end{equation}
Correspondingly, from the point of view of the vacuum structure of the theory, $C^{(W)}$ can be replaced by an ``elementary"  
$3$-form field $C$.  Correspondingly, the
$\theta$-vacuum structure of the 
$SU(N)$ gauge theory is fully captured by 
a theory of a massless 
$3$-form $C$. This effective description of the vacuum is exact, since high-derivative operators 
obtained by integrating out heavy modes vanish 
in the zero momentum limit~\cite{Dvali:2005an}. 

In the full theory, the kinetic function of the $3$-form represents a non-linear algebraic function of the field strength~$E$,
\begin{align}
    \mathcal{L}=\mathcal{K}(E).
\end{align}
The corresponding equation of motion,
\begin{align}
\label{eq:K}
    \partial_\mu \pdv{\mathcal{K}(E)}{E}=0,
\end{align}
has a solution with an arbitrary constant value
$\partial\mathcal{K}(E)/\partial E = \theta \Lambda^2$
representing the $\theta$-vacua.
By integrating the equation \eqref{eq:K}, one gets
$E=\text{Inv}[\mathcal{K}'](\theta)$, where $\text{Inv}[\mathcal{K}']$ denotes the inverse function of  $\partial\mathcal{K}(E)/\partial E$.

Due to the periodicity of physics in $\theta$, $\text{Inv}[\mathcal{K}']$ must be a periodic function. 
For example, in a dilute instanton gas approximation, this function (in units of the 
scale $\Lambda$) has the following form \cite{Dvali:2005an}, 
   \begin{eqnarray} \label{EcosA}
 {\mathcal K}(E) \, =\, E \arcsin(E) \,
 + \, \sqrt{1-E^2} \,, 
\end{eqnarray}
which correctly reproduces the $V(\theta) \propto  - \cos(\theta)$ form of the energy of the $\theta$-vacuum.  

For the purposes of the present work, we will restrict ourselves to the bilinear form of the kinetic term in the Lagrangian $\mathcal{K}(E)=\frac{1}{2}E^2,$ which correctly reproduces the physics of the $\theta$-vacua for $\theta \ll 1$.

Before moving to the monopole case, let us demonstrate how the $3$-form language captures physical effects of the $\theta$-term at the level of classical equations. 
The classicalization of the Chern-Simons $3$-form in the presence of sources can be made transparent by following the analysis of 
\cite{Dvali:2005zk}, 
where the effect of sourcing of an $SU(N)$ Chern-Simons $3$-form by a classical brane was studied.  
Let us evaluate the (gauge-invariant) force mediated by the Chern-Simons $3$-form field between the two conserved brane sources represented by a conserved current $J^{\alpha\beta\gamma}(x)$,
\begin{equation}
J^{\alpha\beta\gamma}(x)\, C^{(W)}_{\alpha\beta\gamma},
\end{equation}
where $\partial_{\alpha}J^{\alpha\beta\gamma}(x) =0 $. For example,
\begin{equation} \label{Bcurrent}
 J^{\alpha\beta\gamma}(x) = q\int \dd^3\xi\, 
 \delta^{(4)}(x- X(\xi)) 
 \pdv{X^\alpha}{\xi^a}\pdv{X^\beta}{\xi^b}\pdv{X^\gamma}{\xi^c}\varepsilon_{abc} 
\end{equation} 
describes a current of a membrane with $2+1$-dimensional world-volume coordinates $\xi^{a},~a=0,1,2$ and $3+1$-dimensional embedding coordinates $X^{\alpha}, ~\alpha = 0,1,2,3$. Here $q$ is a constant charge density,
which we refer to as the $3$-form charge of the membrane. 

As discussed in~\cite{Dvali:2005zk} (see also~\cite{Dvali:2004tma, Sakhelashvili:2021eid}), the role of the membrane, in particular, can be played by an axion-like domain wall created by a pseudo-scalar field $a(x)$ with a periodic potential, e.g., $V(a) \propto (1-\cos(a))$.
Such a theory allows a domain wall across which the field interpolates between the two neighboring vacua, say $0$ and $2\pi$. The role of the conserved brane current sourcing the $3$-form is played by the topological current of the domain wall: $J^{\alpha\beta\gamma}(x) = q\,  \varepsilon^{\alpha\beta\gamma\mu}\,  \partial_{\mu} a$. 
It is easy to see that in an effective theory at distances larger than the domain wall thickness, the topological current of the axion wall matches the brane current~\eqref{Bcurrent}.
 
For a $2$-brane located at $x_3=L$, the current is $J^{\alpha\beta\gamma}(x) =q\, 
\delta(x_3 -L)\, \delta_{0}^{[\alpha} \delta_{1}^{\beta} \delta_{2}^{\gamma]}$. 
 
Now it is straightforward to see that 
the correlator~\eqref{CCcorr} evaluated between two parallel static membranes $J_1$ an $J_2$
(with charges $q_1$ and $q_2$) located in the planes $x_3 =0$ and $x_3 = L$ respectively, asymptotically ($L \rightarrow \infty$) gives a long-range linear potential: 
\begin{equation}
 V(L) \propto  \int \dd^4x \, J_1(x)\, \langle C^{(W)}(x), C^{W}(0)\rangle\, J_2(0) \,  \propto \, q_1q_2\, L \,.
\end{equation}

This potential obtained in a quantum theory is fully matched by the energy of a classical $3$-form field $C_{012} \propto |x_3|$. The corresponding electric field $E = \partial_3 C_{012}$ is a solution of the classical equations of motion of a massless $3$-form \eqref{eq:K}, which in the presence of the brane sources becomes
\begin{align}
\label{eq:KB2}
    \partial_\mu E \, =\frac{1}{3!} \,  
    \varepsilon_{\mu\alpha\beta\gamma} J^{\alpha\beta\gamma}(x) \,.
\end{align}
   
For a single planar brane of charge $q_1$, located at $x_3=0$, we have $J^{\alpha\beta\gamma}(x) =q\, 
\delta(x_3)\, \delta_{0}^{[\alpha} \delta_{1}^{\beta} \delta_{2}^{\gamma]}$. 
The electric field $E$ is constant away from the brane.  However, across it,  $E$ experiences a jump given by $\Delta E \propto q$.
Correspondingly, a probe membrane with a charge $q_1$, placed in the background $3$-form field produced by another source membrane of charge $q_2$, will experience a force perpendicular to its surface and proportional to $q_1 q_2$. 

Notice that the above force also represents the proof of the gauge-independence of the massless pole in equation~\eqref{CCcorr}~\cite{Dvali:2024zpc}.
   
Now we can apply the above understanding to the effect of the $\theta$-term on a probe membrane with charge $q$. 
First, as in \eqref{Boundary1}, we can rewrite the $\theta$-term as the $3$-form field sourced by the space-time boundary
\begin{equation}
   \int_{3+1}\dd^4x \,  \theta \Tr (G_{\mu\nu}\tilde{G}^{\mu\nu})=
   \frac{\theta \Lambda^2}{3!} \int_{2+1} \dd X^{\mu}\wedge \dd X^{\nu} \wedge \dd X^{\alpha} \, 
   C_{\mu\nu\alpha}^{(W)} \,.
\end{equation}
  
This sourcing produces the space-filling constant 
electric field $E$. This expectation value, which is quantum in origin, from the point of view of a local observer, is a Lorentz scalar and has no preferred direction. 

Now, if we introduce a probe membrane, charged under $C$, the field shall act on it non-trivially. The constant force exerted on the probe brane is $\propto q\theta$. In other words, the presence of the classical probe brane effectively classicalizes the 
effect of the $\theta$-term via the $3$-form correlator of the TSV. In particular, it creates a preferred direction through the jump in $E$ over the probe brane. 

Of course, in the above example, we have introduced an axion-like heavy field as a probe and restricted ourselves to a bilinear kinetic function of the $3$-form. This is justified for small 
values of $\theta$ and small charges $q$ of the probe brane.  If the axion is introduced as a part of the full theory, the additional consistency requirements on the charge $q$ as well as on the axion potential shall be imposed. 

The bottom line is that the $3$-form formulation of the $\theta$-vacuum dynamics allows us to analyze the effect of the $\theta$-term, which originates from quantum instanton effects, at the level of the classical equations of motion.

\section{Setting the framework for the numerical analysis}
Equipped with the above knowledge, we proceed in the following way. The $\theta$-term has a physical manifestation due to quantum effects generated by instantons. 
The description in terms of a massless $3$-form allows us to account for these effects at the level of the classical equations of motion of the effective theory. 

Hence, we can include the information about the physicality of $\theta$ into our Yang-Mills Lagrangian by considering the presence of the massless $3$-form. This can be done explicitly by adding to the Lagrangian a constraint via a Lagrange multiplier field $X$, 
\begin{align}
\label{eq:lagrangian}
    \mathcal{L}=-\frac{1}{2}\Tr (G_{\mu\nu}G^{\mu\nu}) + \frac{1}{2}E^2- \, X\left(\Lambda^2 E - \Tr(G_{\mu\nu}\tilde{G}^{\mu\nu})\right).
\end{align}

The above Lagrangian describes the simplest effective theory that promotes the expectation value of $G\tilde{G}$, and thus the $\theta$-term, into an integration constant. Therefore, it exactly captures the structure of the $\theta$-vacuum. Since we approximated ${\mathcal K} = \frac{1}{2} E^2$, the correspondence is valid for small $\theta$. For larger values, one has to include the entire kinetic function.

As shown in~\cite{Dvali:2005an}, the description of $\theta$-vacua by a theory of a massless $3$-form is exact because all high-derivative operators (i.e., operators including derivatives of $E$) vanish in the zero-momentum limit and only the terms algebraic in $E$ are important. 

The constraint, at the level of the classical equations of motion, projects the composite  Chern-Simons $3$-form $C^{(W)}$ of the gauge field $W^a_\mu$ on a massless $3$-form $C$.
In particular, this gives the relation (\ref{eq:GGdual-E-relation}).
Integrating out the Lagrange multiplier then gives
\begin{align}
\label{eq:GGtilde-theta-relation}
    X=\frac{1}{\Lambda^4}G\tilde{G}\equiv \theta \frac{g^2}{16\pi^2}.
\end{align}

Notice that for the purpose of the Witten effect, all we need to know is that $\theta$ is in one-to-one correspondence with the vacuum expectation value (VEV) of $G\tilde{G}$. The rest of the analysis is to show that on a background with this VEV, the monopole acquires a non-zero electric charge matching Witten's formula. 

However, the VEV of $G\tilde{G}$ requires some clarification.
In a Higgsed $SU(2)$ theory, this expectation value originates from constrained instantons, which at the same time are responsible for a non-zero TSV. Therefore,  although this VEV can fully legitimately be described in terms of solutions of the equations of motion of a massless $3$-form, it is not directly factorizable in terms of classical electric and magnetic fields of the low-energy $U(1)$ theory. 

Nevertheless, from the point of view of detecting the Witten effect, it is legitimate to  ``manufacture"  the VEV of $G\tilde{G}$ in terms of electric and magnetic fields of the $U(1)$ theory. However, this question is linked with the issue of Poincaré invariance of the vacuum. 

Let us consider it in more detail. 
We use the language of the Abelian $U(1)$ subgroup,  with the Maxwellian field strength tensor $F_{\mu\nu}$, for which we define the electric field, $\vb{E}$, and the magnetic field, $\vb{B}$
in the usual way. 

As a representative of a classical background with non-zero $F\tilde{F}$, let us choose the configuration in which the electric and magnetic fields are constant in space and time, and are equal $\vb{E} = \vb{B}$. Their absolute values are given by a constant 
\begin{equation} \label{EBandV}
 |\vb{E}| = |\vb{B}| = V.   
\end{equation}
Notice that in the absence of any mobile charges, such a state obeys the superselection rule and thereby represents a legitimate vacuum state.  
    
For definiteness, we can orient the fields in $z$-direction, $E_{i} = B_{i} = \delta_{i3} V$. The corresponding energy momentum tensor is given by $T_{\mu\nu} = \text{diag}( 1, 1, 1, -1) V^2$. Notice that the configuration will give $F_{\mu\nu}\tilde{F}^{\mu\nu} = -4V^2$ and $F_{\mu\nu} F^{\mu\nu} = 0$. 

The above field configuration breaks the Poincaré symmetry. Ordinarily, the invariant vacuum state would be obtained by averaging over all group transformations broken by the state. That is, in quantum theory, the state \eqref{EBandV} must be viewed as a coherent state $\ket{E,B}$ over which the expectation values of quantum operators $\hat{\vb{E}}$ and $\hat{\vb{B}}$ are given by \eqref{EBandV}. The invariant vacuum state is obtained by averaging $\ket{\vb{E},\vb{B}}$ over all possible Poincaré transformations: 
\begin{equation} \label{Pinv}
 \ket{\theta}_{\rm inv} = \int_{\text{Poincaré}} \, \ket{\vb{E},\vb{B}} \,.  
\end{equation}
 This state serves as a qualitative prototype of the invariant $\theta$-vacuum viewed from the point of view of low-energy electric and magnetic fields.  

Of course, in the full $SU(2)$ theory, the expectation value of $G\tilde{G}$ is generated by short-distance physics of constrained instantons, which is not simply reducible to electric and magnetic fields of the effective $U(1)$ theory. However, thinking of the $\theta$-vacua  $\ket{\theta}_{\rm inv}$ in terms of such configurations provides a useful intuition for understanding the  Witten effect in the $3$-form language.

Now, the presence of a magnetic monopole on the above background has two effects, which must be clearly separated. First, the monopole ``classicalizes" the state. The meaning of this is that a classical configuration of the monopole affects the Poincaré invariance of the quantum superposition by creating a preferred frame of reference. In other words, it gives a preference to certain states $\ket{E,B}$ which otherwise would enter the superposition democratically. 

Secondly, the monopole is going to experience the induced charge in each state, regardless of the classicalization.  Although it is important not to confuse the two effects, they go hand in hand.
 
Our analysis indicates that the induced electric charge only depends on the value of $G\tilde{G}$. Since this quantity is the same in all the states obtained by Poincaré transformations, it appears that the Witten effect is insensitive to classicalization and can be considered independently.

\section{Higgsed $SU(2)$ Model}
\label{sec:thooft-polyakov-magnetic-monopole}
We turn now to an $SU(2)$ gauge theory with an adjoint scalar field $\phi$. 
We wish to study the influence of a non-trivial $G\tilde{G}$ background on a monopole.
The Lagrangian of the theory is given by
\begin{align}
    \mathcal{L}=&-\frac{1}{2}\Tr(G_{\mu\nu}G^{\mu\nu})+\Tr((D_\mu \phi)^\dagger (D^\mu \phi))-V_\phi(\phi).
\end{align}

The covariant derivative is defined as usual by $D_\mu \phi =\partial_\mu \phi -i g [W_\mu, \phi]$. The field equations are
\begin{align}
\label{eq:field-equations}
\begin{split}
    &(D_\mu G^{\mu\nu})^a+g \varepsilon_{abc}\phi^b (D^\nu \phi)^c=0,\\ 
    &(D_\mu D^\mu \phi)^a +\pdv{V_\phi}{\phi^a}=0,
\end{split}
\end{align}
where the $SU(2)$ components of the fields are defined by $W_\mu = W^a_\mu T^a$ and $\phi=\phi^a T^a$. The generators are normalized by $\Tr( T_a T_b)=\delta_{ab}/2$.
The potential has the following form
\begin{align}
    V_\phi(\phi)=\lambda \left(\Tr \phi^2-\frac{v^2}{2}\right)^2.
\end{align}
The potential allows the existence of magnetic monopoles, as it breaks the $SU(2)$ gauge symmetry down to $U(1)$. The 't Hooft-Polyakov magnetic monopole solution has the form~\cite{tHooft:1974kcl, Polyakov:1974ek}
\begin{align}
\label{eq:tHooft-Polyakov-magnetic-monopole}
\begin{split}
    \phi^a&=\frac{1}{g}\frac{r^a}{r^2}H(r),\\
    W^a_i&=\frac{1}{g}\varepsilon_{aij}\frac{r^j}{r^2}(1-K(r)),\\
    W^a_t&=0,
    \end{split}
\end{align}
where $H(r)$ and $K(r)$ are profile functions that depend on the parameter $m_h/m_v$, with the Higgs boson mass $m_h=\sqrt{2\lambda} v$ and the vector boson mass $m_v=gv$. 

In~\cite{tHooft:1974kcl}, 't Hooft found  that the gauge invariant electromagnetic field strength tensor can be written as
\begin{align}
    \label{eq:tHooft-field-strength}
    F_{\mu\nu}=\hat{\phi}^a G_{\mu\nu}^a-\frac{1}{g}\varepsilon_{abc}\hat{\phi}^a (D_\mu \hat{\phi})^b (D_\nu \hat{\phi})^c,
\end{align}
(see also~\cite{Rajaraman:1982is,Weinberg:2012pjx}).
Inserting ansatz \eqref{eq:tHooft-Polyakov-magnetic-monopole} into \eqref{eq:tHooft-field-strength} gives the magnetic field
\begin{align}
    B_i=-\frac{1}{2}\varepsilon_{ijk}F_{jk} \xrightarrow{r \rightarrow \infty} \frac{1}{g}\frac{r_i}{r^3},
\end{align}
describing a magnetic monopole with charge $q_m=4\pi/g$.  

A background with a constant electric and magnetic field in the $z$-direction is given by
\begin{align}
    W^\text{background}_{a,\mu}=\hat{n}^a \begin{pmatrix}
        |\vb{E}| z\\
        0\\
        |\vb{B}| x\\
        0
    \end{pmatrix}.
\end{align}
On top of this, we can set the magnetic monopole with
\begin{align}
    \label{eq:ansatz-monopole-constant-GGdual}W_{a,\mu}=W^\text{background}_{a,\mu}+W^\text{monopole}_{a,\mu}.
\end{align}
This ansatz is not a solution to the static field equations; however, it serves as a good initial configuration to find a solution by numerical relaxation.
Notice that by using Dirichlet boundary conditions, the constraint~\eqref{eq:GGdual-E-relation} stays automatically satisfied at the boundary.

\section{Numerical Details}
\label{sec:numerical-details}
For the relaxation, we used the iteration
\begin{align} 
    f^{(n+1)} = \frac{\delta^2}{6} E[f^{(n)}] + f^{(n)}, 
\end{align}  
where $f$ is the field being relaxed, $\delta$ is the lattice spacing, and $E[f] = 0$ represents its static equation of motion.  
This iterative procedure updates  $f^{(n)}$ by adding a correction proportional to the residual error in satisfying $E[f] = 0$. Once $f^{(n)}$ converges such that $E[f^{(n)}] = 0$, the solution is obtained. A more detailed explanation of this method can be found in~\cite{Bachmaier:2025jaz, Patel:2023sfm}.

In our scenario, we relaxed $W^a_i$ and $\abs{\phi} = \sqrt{\phi^a \phi^a}$, while keeping the direction of the scalar field $\hat{\phi}^a$ fixed throughout the entire relaxation process. Consequently, $\partial_i \hat{\phi}^a$ and $\partial_i^2 \hat{\phi}^a$ also remain unchanged.  

For the computation, we used the programming language Python together with the Numba package~\cite{Numba}, which translates the Python code into fast machine code and provides a simple implementation to parallelize the code to take advantage of multi-core processors. 
We performed the computation on a cubic lattice with $180^3$ lattice points and a lattice spacing of $0.5 m_v^{-1}$. As a cross-check, we used for a few cases a lattice spacing of $0.25 m_v^{-1}$. Furthermore, we tested different box sizes to assess potential boundary effects. We found that these variations lead to negligible changes in our results. 

At each iteration step of the relaxation process, we measured the electric charge within a finite region around the center. Once the charge converged to a stable value (relative change between iteration steps below $10^{-6}$) we terminated the relaxation. This occurred before $10.000$ iteration steps.
The total energy also converged after this number of iterations.
For cases with only an electric field in the background, the total charge of the system is always zero, and thus, we performed $10.000$ iterations without stopping the iteration process before. After relaxation, we also verified that the Gauss constraint was satisfied. 

\section{Polarizability of the monopole}
\label{sec:polarizability-of-the-magnetic-monopole}
The first step towards a constant $G\tilde{G}$ background is to consider a monopole in an electric field without a background magnetic field.
Already there, we can observe non-linear effects, most prominently the polarizability of the monopole.

We would like to emphasize that, 
since on such a background $G\tilde{G} = 0$, the phenomenon is unrelated to the Witten effect.
However, the polarizability of the monopole is interesting on its own right, regardless of our use of it as an intermediate step in deriving the Witten effect. We are not aware of an explicit discussion of the polarizability of a 't Hooft-Polyakov magnetic monopole in a background electric field in a pure bosonic theory. Therefore, we devote a separate discussion to this effect. 

One can understand this phenomenon in the particle physics picture in which the monopole is viewed as a composite object, a coherent bound state, of $W$ and scalar bosons~\cite{Dvali:2011aa, Dvali:2019jjw}. 
The following features are shared with  generic topological solitons \cite{Dvali:2015jxa, Muck:2015dea, Berezhiani:2024pub}. 
The constituents can be split into two categories. The ones that predominantly contribute 
to the energy have wavelengths comparable to the size of the monopole core and are localized within it. 
The constituents that are responsible for the topological charge of the soliton, have infinite wavelengths as well as infinite occupation numbers 
~\cite{Dvali:2015jxa, Berezhiani:2024pub}. These shall be considered later. The polarizability 
of the monopole is mainly due to the energy constituents, which have the following properties. 

The occupation number of gauge $W$-bosons is given by $N_v \sim 1/g^2$, and this matches the number of Goldstones, which represent their longitudinal polarizations. The occupation number of Higgs bosons 
(the radial mode) is correspondingly given by the inverse Higgs coupling. 
 
Of course, the constituent $W$ and scalar bosons are off-shell as compared to their asymptotic counterparts, and they are stabilized by the topological charge. Despite being 
off-shell, one can consistently talk about their 
characteristics, such as the occupation number, the wavelength, and the electric charge.
Although the $W$ and scalar bosons are electrically charged, the occupation numbers of the opposite charges in the ground-state (assuming a $\theta=0$ vacuum) are equal. 
Due to this, the monopole is electrically neutral. 
However, under the influence of an external electric field, the oppositely charged bosons can separate and generate a dipole moment.

The polarizability of the monopole is also evident from the equations of motion. Taking the zero component of the gauge field equation~\eqref{eq:field-equations} and contracting it with $\hat{\phi}^a$ projects out the unbroken $U(1)$ direction. The resulting equation can be rewritten as
\begin{align} 
    \label{eq:Gauss-law}
    \partial_i (G^a_{0i}\hat{\phi}^a) = G^a_{0i} (D_i \hat{\phi})^a. 
\end{align}
The left-hand side represents the divergence of the electric field, implying that the right-hand side corresponds to the charge density. Taking $\abs{\vb{B}}=0$ in the ansatz~\eqref{eq:ansatz-monopole-constant-GGdual} and inserting it into~\eqref{eq:Gauss-law} gives
\begin{align} 
    \vb{\nabla} \cdot \vb{E} = 2|\vb{E}| \frac{z}{r^2}K(r)^2,
\end{align}
which explicitly shows that the charge flips sign at $z = 0$.

Applying the relaxation method gives an electric dipole field
\begin{align}
\label{eq:dipole-electric-field}
    \vb{E}^\text{dipole} =\frac{3 (\vb{p}\cdot \vb{\hat{n}}) \vb{\hat{n}}-\vb{p}}{r^3},
\end{align}
on top of the background.
Our numerical analysis showed that the dipole moment, $\vb{p}$, is non-zero and is linearly proportional to the background electric field, indicating that the magnetic monopole is polarizable:
\begin{align}
    |\vb{p}|=P\, |\vb{E}|,
\end{align}
where $P$ is the polarizability of the magnetic monopole.

\begin{figure}[t]
    \includegraphics[trim=0 0 0 30,clip,width=\linewidth]{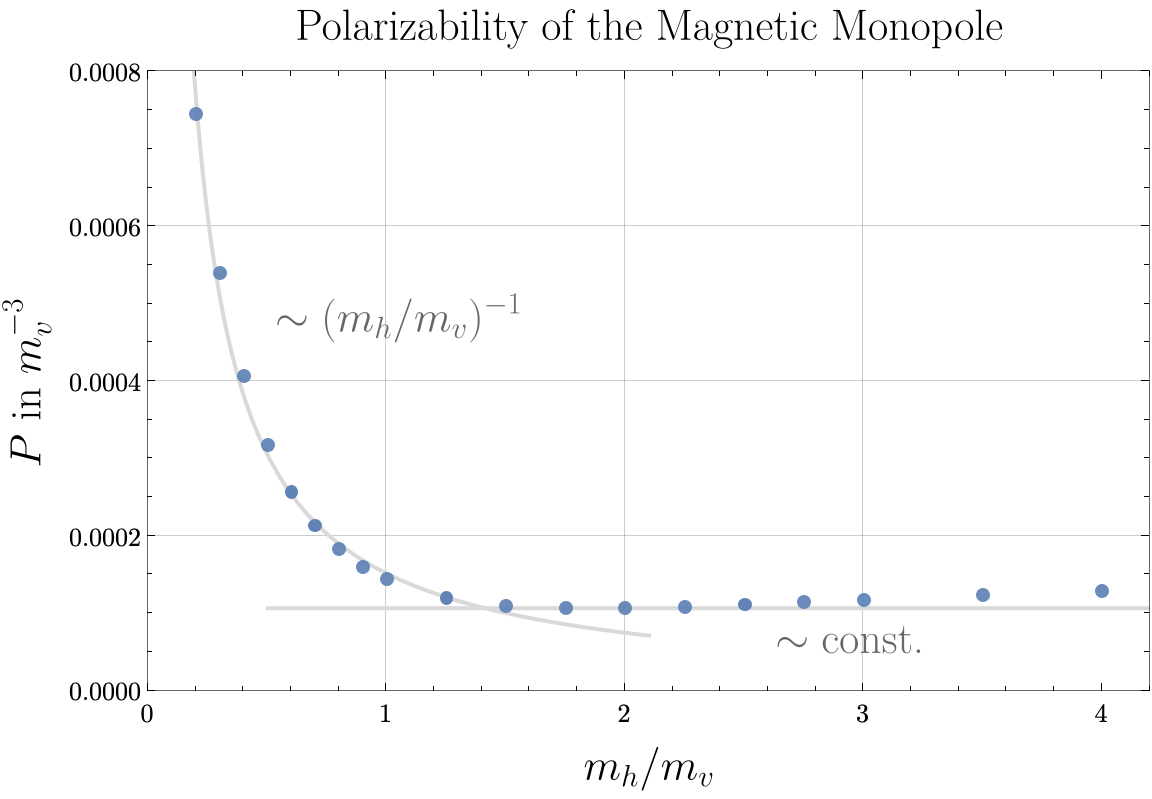}
    \caption{Polarizability of the monopole with respect to $m_h/m_v$. For $m_h \lesssim m_v$, the Polarizability decays approximately with $(m_h/m_v)^{-1}$, and for $m_h \gtrsim m_v$, it stays approximately constant.}
    \label{fig:polarizability}
\end{figure}

Using the numerical relaxation method described above, we obtained an electric dipole field superimposed on the background electric field. By comparing the numerical results with the expected dipole field behavior in equation~\eqref{eq:dipole-electric-field}, we determined the dipole moment. As expected, it is linearly proportional to the background electric field. 
The calculated polarizability for different $m_h/m_v$ is illustrated in FIG.\ref{fig:polarizability}.

For $m_h\lesssim m_v$, the polarizability decreases with $(m_h/m_v)^{-1}$. 
Our lattice does not capture the long-range electric field behavior for values $m_h/m_v<0.2$, because the monopole core size approaches the lattice size. For $m_h\gtrsim m_v$, the polarizability stays approximately constant.

Our interpretation of the numerical results is the following. 
We can approximate the dipole by two opposite charges, $\pm q$, separated by a distance $d$. The maximal available charge $q$ is given by $N_{v} g$, where $N_v \sim 1/g^2$ is 
the occupation number of $W$-bosons. The maximal length $d$ is given by the largest of the two cores (Higgs versus gauge). Notice that the occupation numbers of the $W$-bosons must be the same as the ones 
of the charged Higgs components since the latter represent the longitudinal modes of the massive $W$-bosons. However, because the monopole background breaks Poincaré invariance and also partially ``restores" the $SU(2)$ symmetry in the core, the spread of the wave-function profiles for the $W$ and Goldstone bosons can differ.
Correspondingly, in different regimes, $d$ is expected to be given by 
\begin{align}
    d\sim
    \begin{cases}
        \frac{1}{m_h}&\text{for}\, \,  m_h\lesssim m_v\\
        \frac{1}{m_v}&\text{for}\, \,  m_h\gtrsim m_v\, .
    \end{cases}
\end{align}

At the same time, the value of the electric field is bounded from above by the value $\abs{\vb{E}_\text{max}}$ for which its energy within the monopole core becomes equal to the monopole mass, $\abs{\vb{E}_\text{max}}^2 m_{v}^{-3} = m_{v}/g^2$. Notice that this gives the bound on the electric field given by the Schwinger threshold, $\abs{\vb{E}_\text{max}} =  m_{v}^2/g$. 
Therefore, the polarizability is
\begin{align}
\label{eq:polarizability-estimation}
    P\propto \frac{N_v g\, d}{|\vb{E}_\text{max}|}\sim
    \begin{cases}
        \frac{1}{m_v^2}\frac{1}{m_h}&\text{for}\, \,  m_h\lesssim m_v\\
        \frac{1}{m_v^3}&\text{for}\, \,  m_h\gtrsim m_v\, .
    \end{cases}
\end{align}

Our expectation from equation~\eqref{eq:polarizability-estimation} is that $P$ stays constant for $m_h\gtrsim m_v$. However, the precise monopole mass is~$M_\text{monopole}=f(m_h/m_v)\, m_v/g^2$, where $f(m_h/m_v)$ is a monotonically increasing function from $f(m_h/m_v\rightarrow 0)=1$ to $f(m_h/m_v\rightarrow \infty)\approx1.787$~\cite{Forgacs:2005vx}. Therefore, the occupation number of $W$-bosons increases slightly with $N_v\sim f(m_h/m_v)/g^2$, which may explain the small increase in the observed polarizability.

\section{Witten Effect}
\label{sec:Witten-effect}
Now, in addition to the electric field background, we also turn on a magnetic field, leading to a constant $G\tilde{G}$ background. We set $|\vb{E}|=0.05 m_v^2$ and increased the magnetic field by $\Delta|\vb{B}| =0.005 m_v^2$ steps.
Numerically, we always used the previously relaxed configuration, updated the magnetic field background, and then applied the relaxation again to obtain the new result.

\begin{figure}[t]
    \includegraphics[trim=0 0 0 30,clip,width=\linewidth]{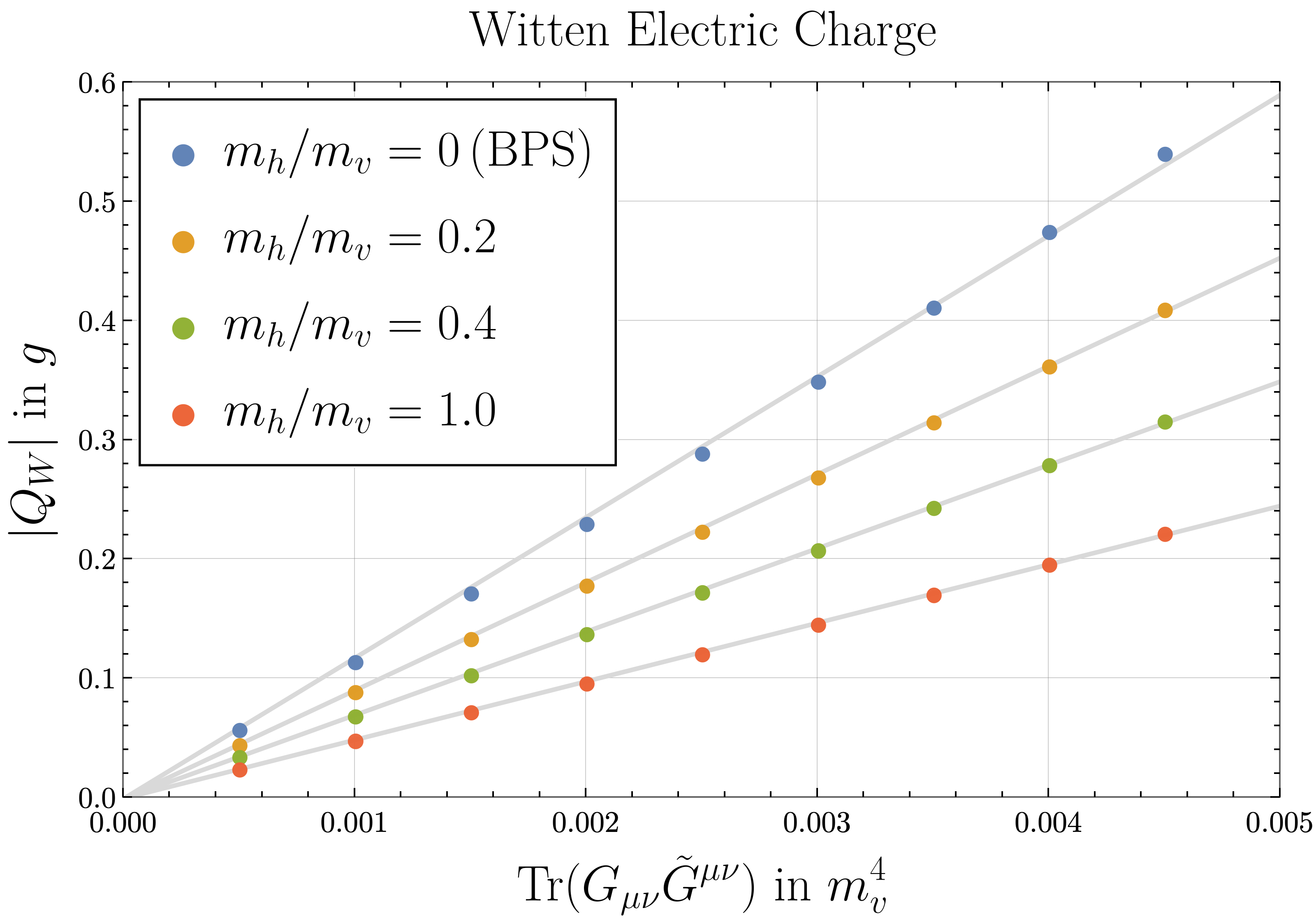}
    \caption{Witten electric charge with respect to $G\tilde{G}$ for different values of $m_h/m_v$. The electric charge is directly proportional to $G\tilde{G}$ as can be seen from the linear fit (gray lines).}
    \label{fig:wittenelectriccharge}
\end{figure}

At each step, we calculated the induced electric charge $Q_W$ by evaluating the electric flux through a cubic box with edge length $l=45 m_v^{-1}$ (half of the lattice size). 
Outside of the box, the electric field on top of the background already decays as $r^{-2}$, meaning that all the charge is within this region.
The results of this analysis are shown in FIG.~\ref{fig:wittenelectriccharge}.

We can observe that the electric charge is linearly proportional to $G\tilde{G}$. 
We numerically analyzed various Poincaré-transformed configurations of the background electric and magnetic fields. Our analysis showed that the induced Witten electric charge remains invariant under these transformations. This result is 
fully consistent with the expectation that with all other invariants of the background vanishing, the
induced charge depends solely on $G\tilde{G}$.

As explained in section~\ref{sec:3-form-formulation-of-the-theta-vacuum}, this constant $G\tilde{G}$ background can be related to $\theta$. Therefore, the appearance of the electric charge can be interpreted as the Witten effect. 

This phenomenon appears in the pure 't Hooft-Polyakov model described in section~\ref{sec:thooft-polyakov-magnetic-monopole}, 
since the information about the $\theta$-vacuum
is encoded in the VEV of $G\tilde{G}$, which represents the field strength of a massless 
$3$-form field. 
Non-linear effects allow the background field to interact with the magnetic monopole, leading to the Witten effect. 
This is in contrast to the Dirac monopole of the $U(1)$ theory, as we will discuss in the appendix.

FIG.~\ref{fig:wittenelectricchargemh} shows the dependence of the Witten charge on $m_h/m_v$. We found that the electric charge approaches a constant value that is proportional to $G\tilde{G}$ for $m_h/m_v\rightarrow \infty$. This value is reached with an $(m_h/m_v)^{-1}$ behavior. The maximal Witten charge occurs in the BPS case~$m_h/m_v=0$.

\begin{figure}[t]
    \includegraphics[trim=0 0 0 25,clip,width=\linewidth]{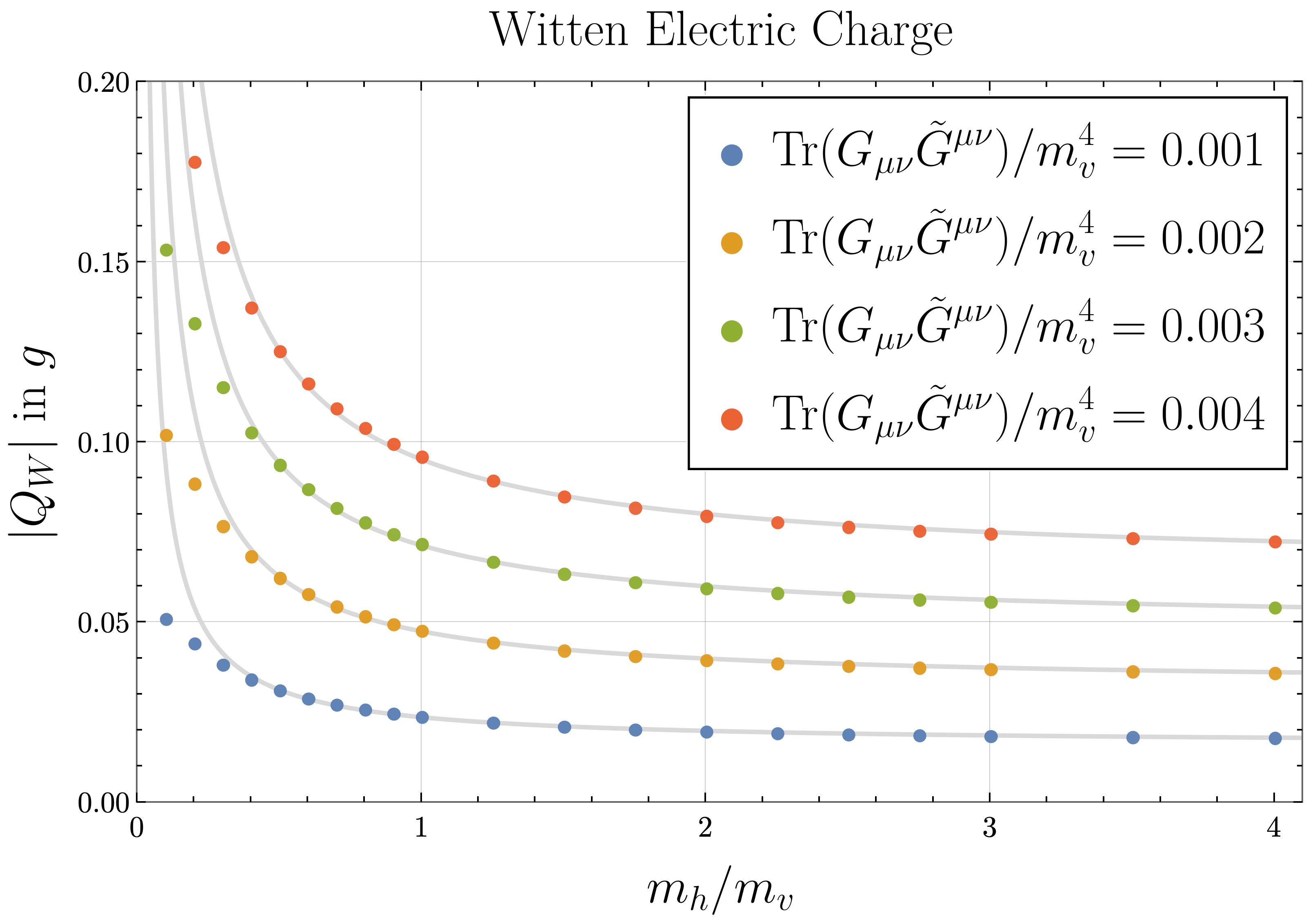}
    \caption{Witten charge with respect to $m_h/m_v$ for different $G\tilde{G}$. The electric charge approaches a constant value for large $m_h$ with a $(m_h/m_v)^{-1}$ behavior (gray lines).}
    \label{fig:wittenelectricchargemh}
\end{figure} 

The described parameter dependence of the Witten charge must not create an impression that this charge can violate equation~\eqref{eq:Witten-charge}. The point is that the relation \eqref{eq:GGtilde-theta-relation} between the expectation value of $G\tilde{G}$ and the $\theta$-term is determined by the scale $\Lambda$, which, of course, depends on the parameters of the theory non-trivially. In particular, 
this dependence emerges through the 
parameter dependence of the contribution of constrained instantons.
 
For a fixed value of $G\tilde{G}$, the $m_h/m_v$-dependent Witten electric charge can be understood as follows. Within the monopole, $W$-bosons interact both via the gauge and the scalar forces. For $m_h \ll m_v$, the Higgs-mediated interactions are present everywhere within the monopole's gauge core, allowing more $W$-bosons to be bound together and thus the electric charge can be very large. In contrast, for $m_h\gg m_v$, the Higgs interactions are strongly suppressed, leaving the gauge interactions to dominate.
Since these gauge interactions maintain the same order strength for fixed $m_v$, the number of bound $W$-bosons approaches a constant value.
Correspondingly, so does the electric charge.\footnote{The same argument can be applied to the maximal electric charge of a Julia-Zee dyon~\cite{BachmaierDyonCharge}.}

In our numerical analysis, we observed that in the limit of small Higgs boson masses, the Witten charge is not strictly linearly proportional to $G\tilde{G}$. Instead, the charge increases slightly more strongly than a purely linear dependence would suggest.
This is due to the fact that our analysis is valid only for small values of $\theta$, which was reflected in the choice of  $\mathcal{K}(E)$ as bilinear. This reproduces correctly the $\theta$-dependence of $G\tilde{G}\propto\theta$ for a small argument, which is of course linear.
However, for larger values of $\theta$, one has to take the full non-linear $\mathcal{K}$ which makes $G\tilde{G}$ non-linear in $\theta$. For example, in the dilute instanton gas approximation, it has the form~\eqref{EcosA} which gives $G\tilde{G}\propto \sin(\theta)$. Therefore, for the Witten electric charge we expect an $\arcsin(G\tilde{G})$-like behavior for larger values of $G\tilde{G}$. 
A more detailed analysis of these higher-order effects is beyond the scope of our work.

\section{Expected effects of virtual monopoles}
\label{sec:expected-effects-of-virtual-monopoles}

The $3$-form TSV correlator picture also answers the question of why we do not see an ``inverse Witten effect" for the electric charges.
In other words, it provides a particle physics explanation for why the electric quanta do not acquire small magnetic charges in a $\theta$-vacuum.  

Of course, this question may be viewed as groundless, since in the weakly coupled theory there is a fundamental asymmetry between magnetic and electric charges. For instance, the electric field strength that would match the strength of the 
magnetic one of the monopole, reaches the electric Schwinger threshold. At the same time, the magnetic field is way below the one 
required for a would-be magnetic Schwinger effect. Despite these obvious differences, it is useful to understand things from other perspectives. 
 
Unlike the electrically charged elementary particles, the magnetic charges are represented either by infinitely-heavy and singular objects (as in the case of the Dirac monopole in the $U(1)$ theory) or as solitons with topological charge (as in the case 
of the 't Hooft-Polyakov monopole in the $SU(2)$ theory). 

From the point of view of the elementary quanta,
any topological soliton, and in particular the 't Hooft-Polyakov monopole, represents a coherent state with a divergent occupation number towards the infinite wavelength of the constituents~\cite{Dvali:2015jxa, Berezhiani:2024pub}. 
As already discussed, in this picture, a monopole
represents a composite object of elementary constituents that can be split into two categories: those that determine the mass and those that determine the topological charge. 

The energy constituents of the soliton have characteristic wavelengths given by the size of the core, and their occupation number is given by the inverse coupling. In contrast, the constituents responsible for the topological charge of the soliton
have divergent occupation numbers towards the infinite wavelengths~\cite{Dvali:2015jxa, Berezhiani:2024pub}. That is, the topological charge of a soliton maps to the singularity in the coherent state parameter in the infrared wavelengths. An explicit BRST invariant construction of such coherent states can be found in~\cite{Berezhiani:2024pub}.

Notice that the infinite occupation number of the constituents explains the stability of the topological soliton, since the decay into a topologically trivial vacuum state would require an infinite shift in the occupation number, which has zero probability. 
In short, the topological solitons are composites with an infinite number of infrared elementary constituents. 
  
This fundamental difference between the monopoles and elementary quanta is the main reason why monopoles can classicalize the $G\tilde{G}$ background, whereas particles cannot. 

Nevertheless, the knowledge of the Witten effect can suggest the presence of certain magnetic effects among the elementary particles. 
To see this, let us think of a monopole-antimonopole pair separated by some finite distance $L$. This system carries zero topological charge. Correspondingly, viewed as a coherent state, such a configuration does not contain an infinite occupation number of IR quanta. The occupation number shall diverge only for $L \rightarrow \infty$. Due to the Witten effect, such a magnetic dipole must produce an electric dipole. 

Let us now compare this system with a pair of elementary quanta of opposite electric charges separated by a distance $L$, creating a dipole moment $\vb{p}_e$.
At finite distances, $L \gg m_v^{-1}$, both systems contain a finite occupation number of quanta. The number of quanta in magnetic monopoles is much larger, $N \sim \frac{(m_v L)^2}{g^2}$, but it is still finite. We do expect that in such a case, a magnetic dipole behavior must be exhibited also by the pair of electric charges. 
Of course, the effect must be exponentially suppressed. 

\begin{figure}[t]
    \centering
    \includegraphics[trim=0 20 0 20,clip,width=0.79\linewidth]{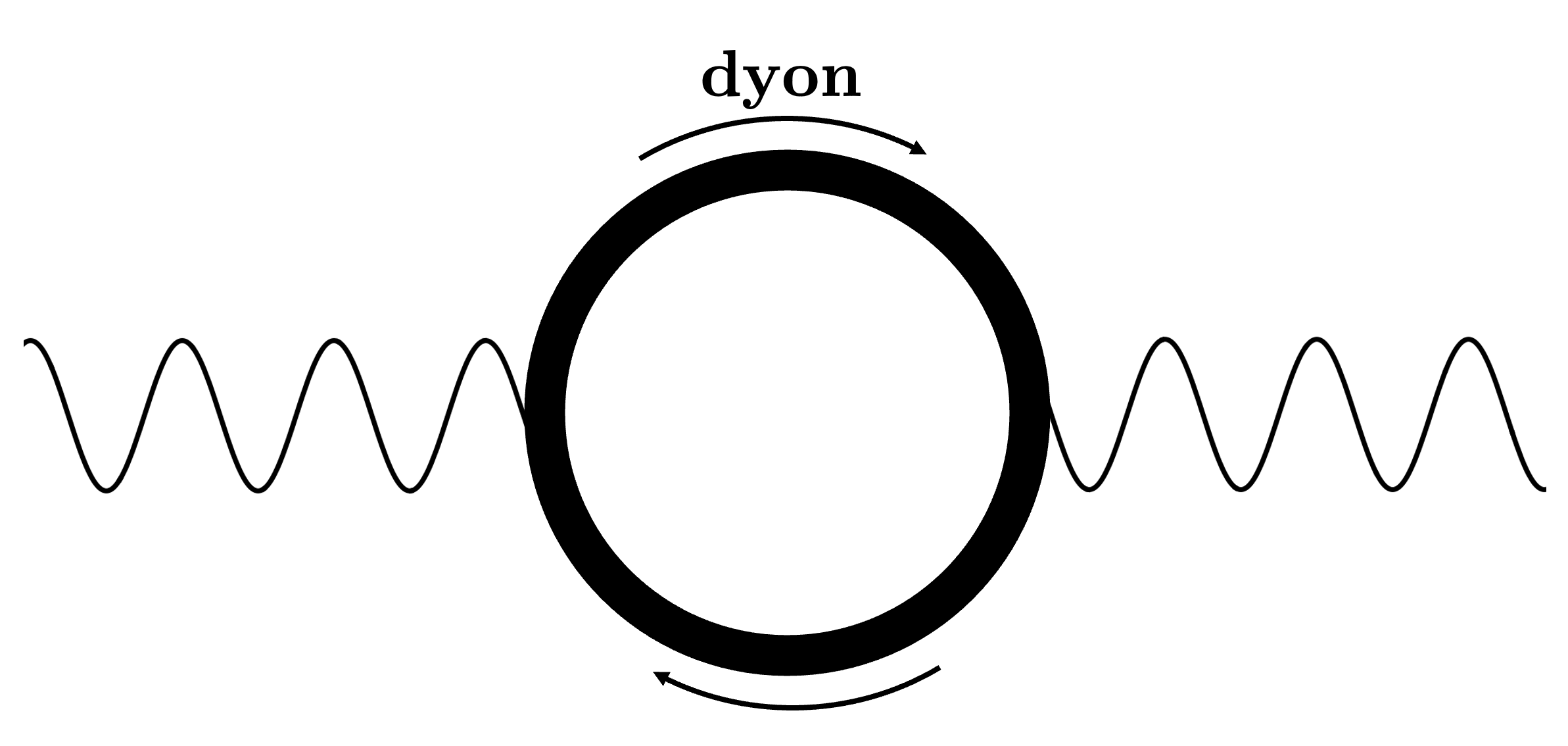}
    \caption{A background electric field (for example, coming from an electric dipole) can polarize the vacuum through virtual dyon-antidyon pairs of electric charge $\pm \theta g$ and magnetic charge $\pm g^{-1}$.}
    \label{fig:dyonloop}
\end{figure}

Indeed, there exists such a candidate vacuum polarization effect coming from the exchange of virtual dyon-antidyon pairs among the elementary charges (see Fig. \ref{fig:dyonloop}). 
We shall not evaluate this contribution explicitly, but the expected suppression can be estimated from the following simple argument.  
The production of an intermediate (virtual) multi-particle state is suppressed by~\cite{Dvali:2020wqi}, 
\begin{equation} 
    e^{-N}\sim e^{-c\frac{(m_v L)^2}{g^2}},
\end{equation}
where $c$ is a numerical factor. 
Thus, we have the following picture: Although an isolated elementary particle carries no magnetic charge, the pair of particles separated at a finite distance carries a magnetic dipole with the dipole moment
\begin{equation} 
\label{eq:Witten-effect-dipole}
 \vb{p}_m \sim \theta g^2\, \vb{p}_e \,  e^{-c\frac{(m_v L)^2}{g^2}} .  
\end{equation}
This effect is exponentially suppressed,
modulo the coefficient, which we did not compute. 
However, on physical grounds, it is expected to be non-zero. Notice that in extended theories in which the monopole can exhibit a near-maximal microstate degeneracy \cite{Dvali:2019jjw}, the effect can be exponentially enhanced by a
degeneracy factor \cite{Dvali:2020wqi}. This can lead to new non-perturbative effects that shall not be discussed here.

\section{Conclusion}
\label{sec:conclusion}

In this work, we studied the Witten effect in the $3$-form description of $\theta$-vacua. In this picture, the $\theta$-vacua represent 
vacua with different VEVs of $G\tilde{G}$. Correspondingly, the Witten effect is obtained as the influence of this background 
on the magnetic monopole.  

Using numerical relaxation, we observed that the magnetic monopole in the presence of non-zero $G\tilde{G}$ indeed acquires an electric charge as predicted by Witten~\cite{Witten:1979ey}.
The quantum nature of the effect is captured by the non-zero VEV of $G\tilde{G}$, which originates from instantons that are responsible for the
TSV  in the $SU(2)$ theory. 

As a byproduct of our analysis, we found that a 't Hooft-Polyakov magnetic monopole undergoes polarization in the presence of a constant electric field. This effect can be understood as the separation of constituent $W$-bosons of opposite charges within the monopole’s core. Our numerical results match the behavior obtained from the picture of a monopole as a composite coherent state of gauge and scalar bosons.

We also suggested that, due to virtual dyon-antidyon pairs, separated charged particles are expected to acquire a magnetic dipole moment proportional to the electric one and $\theta$ (see equation~\eqref{eq:Witten-effect-dipole}). 

As an auxiliary exercise (see appendix), we also analyzed the case of a Dirac monopole in the $U(1)$ theory, which represents the low-energy limit of the 't Hooft-Polyakov model.  However, for a fully transparent $3$-form description of the Witten effect, the UV-completion of the $U(1)$ theory into $SU(2)$ is required, because the 
vacua are made physical by $SU(2)$ instantons.

\section*{Acknowledgements}
We want to thank Josef Seitz for valuable discussions and collaboration at the early stages of this work. We are grateful to Lasha Berezhiani,
Giacomo Contri, Oriol Pujolas, Otari Sakhelashvili, and Tanmay Vachaspati, for useful discussions and helpful comments.

This work was supported in part by the Humboldt Foundation under the Humboldt Professorship Award, by the European Research Council Gravities Horizon Grant AO number: 850 173-6,  by the Deutsche Forschungsgemeinschaft (DFG, German Research Foundation) under Germany's Excellence Strategy - EXC-2111 - 390814868, and Germany's Excellence Strategy under Excellence Cluster Origins.

J.S.V.B. acknowledges support from the Spanish Ministry of Science and Innovation (MICINN) through the Spanish State Research Agency under the R\&D\&I project PID2023-146686NB-C31 funded by MICIU/AEI/10.13039/501100011033/ and by ERDF/EU, and under Severo Ochoa Centres of Excellence Programme 2025-2029 (CEX2024001442-S). The CERCA program of the Generalitat de Catalunya partially funds IFAE. MICIIN supported this study with funding from the European Union NextGenerationEU(PRTR-C17.I1) and by the Generalitat de Catalunya.

\noindent\textit{Disclaimer:}
Funded by the European Union. Views and opinions expressed are, however, those of the authors only and do not necessarily reflect those of the European Union or European Research Council. Neither the European Union nor the granting authority can be held responsible for them.

\section*{Appendix: Witten Effect in $U(1)$  Theory}
In this appendix, we study the effects of the $\theta$-term on a Dirac magnetic monopole of the $U(1)$ theory.
From first glance, this may sound irrelevant, since naively in the $U(1)$ theory, the $\theta$-term is unphysical due to the lack of the topological structure of the vacuum.  However, the physicality of $\theta$ hides in UV-completion, which 
is described by the embedding  
of $U(1)$ into an $SU(2)$-Higgsed theory at some high energy scale. Such an embedding resolves the Dirac monopole of the $U(1)$ theory as the 
't~Hooft-Polyakov monopole of the $SU(2)$ theory. 

The $U(1)$ theory is obtained as the limit in which the masses of gauge and Higgs bosons are taken to infinity.  In this limit, the theory becomes $U(1)$ and at the same time the 't Hooft-Polyakov monopole becomes a Dirac monopole.  On the other hand,  the Witten charge is independent of the scale of the $SU(2)$ theory. Thus, somehow the effect must survive in $U(1)$ theory with the Dirac monopole. 

The above implies that a low energy observer of the $U(1)$ theory must be able to discover the Witten effect (at least qualitatively) under the knowledge of certain features of the UV theory that are able to penetrate into the IR theory of $U(1)$. 
 
Such knowledge can come through the VEV of the dual field strength, or equivalently, the knowledge that the Chern-Simons projects on a massless $3$-form. 
  
Let us consider a $U(1)$ theory including the boundary $\theta$-term into the Maxwell Lagrangian
\begin{align}
   \mathcal{L} =&-\frac{1}{4} F_{\mu \nu} F^{\mu \nu} \, +  \, \frac{1}{4} \theta F_{\mu \nu} \tilde{F}^{\mu \nu} \,, 
\end{align}
and evaluate the theory on the background
of a Dirac monopole. 

As it is well-known, the Dirac monopole comes with a singularity price. 
First, the monopole is singular at the origin.  
Second, the vector potential $A_{\mu}$ is not continuously defined on any sphere surrounding the monopole.
For example, we can choose the following gauge in the spherical coordinates, 
\begin{equation} \label{Dirac}
 A_{\phi} = q_m \frac{1 - \cos(\vartheta)}{r \sin(\vartheta)} \,, ~ A_{\vartheta} = A_r =0 \,,  
\end{equation}
corresponding to a radial magnetic field, 
\begin{equation} \label{DiracB}
 \vb{B}_\text{mon} = q_m \frac{\vb{r}}{r^3}\,,   
\end{equation}
with the magnetic charge $q_m$. 
In this gauge, there is a singularity in the form of a Dirac string coinciding with the negative $ z$-axis. 

Likewise, in the Wu-Yang representation \cite{Wu:1975es}, there is a discontinuity between the vector potentials defined in the upper and lower hemispheres, 
\begin{equation} \label{DiracWY}
 A_{\phi}^{U} = q_m \frac{1 - \cos(\vartheta)}{r \sin(\vartheta)} \,, ~~  
 A_{\phi}^{L} = - q_m \frac{1 + \cos(\vartheta)}{r \sin(\vartheta)} \,.    
\end{equation}
In both representations of the monopole, the requirement of unphysicality of the corresponding irregularities of the vector potentials leads to Dirac's charge quantization condition.
Of course, singularities are resolved by the UV-completion of the theory into $SU(2)$, which imposes the charge quantization at the fundamental level. 
 
The Witten effect can be guessed by a low-energy observer using a certain knowledge of this UV completion.  Namely, it suffices to know that UV physics promotes the VEV 
of $F\tilde{F}$ into an integration constant. This is equivalent to having a massless $3$-form. Using this information, the Witten effect can be observed as it was done in the case of $SU(2)$. 
  
The effective Lagrangian can be taken as, 
\begin{align}
   \mathcal{L} =&-\frac{1}{4} F_{\mu \nu} F^{\mu \nu} + \Lambda^4 {\mathcal K(F\tilde{F}/\Lambda^4)} \,.
\end{align}
The equations of motion are, 
\begin{equation}
\label{KKK}
    \partial_{\mu} F^{\mu \nu} = 2\partial_{\mu}\left({\mathcal K'(F\tilde{F}/\Lambda^4)}\Tilde{F}^{\mu \nu}
    \right) \,.
\end{equation}
The theory has a continuum of vacuum solutions
with equal electric and magnetic fields $\vb{E}_\text{vac} = \vb{B}_\text{vac}$. The corresponding vector potential can be chosen as 
\begin{equation}
A^{\text{vac}}_\mu= -V
\begin{pmatrix}
         z\\
    0\\
    x\\
    0
    \end{pmatrix},
\end{equation} 
which gives, $E_i^\text{vac} = B_i^\text{vac} = \delta_{i3} V$. We have $F\tilde{F} =-4 V^2 = \Lambda^2 E$, while all other invariants are zero.  
As usual, in quantum theory, these classical fields should be viewed as expectation values of the corresponding operators over a coherent state $\ket{\vb{E},\vb{B}}$.
  
In a theory without massless fermions, such  configurations mark different superselection sectors which can be  parameterized by, 
\begin{equation} \label{Kprime}
  \theta=2\mathcal K'(F\tilde{F}/\Lambda^4) \,.
\end{equation} 

Of course, as already discussed, the true Poincaré-invariant vacuum states, $\ket{\theta}_{\rm inv}$, represent superpositions of the states $\ket{\vb{E},\vb{B}}$, obtained by summing over all possible rotations and Lorentz boosts, as in equation~\eqref{Pinv}. 
Obviously, the VEV of  $F\tilde{F}$ (and thus of $\theta)$ is the same in all 
Poincaré-transformed states. Therefore, the invariant vacua~\eqref{Pinv} can be parameterized by $F\tilde{F}$ (or $\theta)$. 
 
In the vacuum $\ket{\theta}_{\rm inv}$ the fields are highly entangled. However,  since the important quantity controlling the Witten effect is the 
VEV of $F\tilde{F}$, we can analyse the effect for a single member $\ket{\vb{E},\vb{B}}$
of the superposition (\ref{Pinv}), assuming  that 
in the Poincaré-averaged vacuum $\ket{\theta}_{\rm inv}$ it remains unchanged. 

Hence, we can place the Dirac monopole on top of such a background.  Then, we can show that it sources the electric flux through an asymptotic sphere  
($r \rightarrow \infty $) enclosing the monopole.
For a static configuration, the time-component of equation~\eqref{KKK} is 
\begin{equation} 
\label{GKKK}
\partial_i E_i \, = \,   2 \partial_i \left({\mathcal K'(F\tilde{F}/\Lambda^4)} \, B_i \right )\,.
\end{equation}  
Again, we assume that there exists UV physics that consistently resolves the singularity at the origin without jeopardizing the validity of the surface integral.  
 
Then, sufficiently far away from the monopole the fields can be split into the vacuum and monopole components as $\vb{E}= \vb{E}_\text{vac} + \vb{E}_\text{mon}$ and $\vb{B} = \vb{B}_\text{vac} + \vb{B}_\text{mon}$. 
Applying the Gauss law, we represent this as an integral over a boundary sphere at 
$r \rightarrow 
\infty$. Taking into account that for large $r$, $\vb{B}_\text{mon} \rightarrow q_m \vb{r}/r^3$ and $\vb{E}_\text{mon} \rightarrow q_e \vb{r}/r^3$, and correspondingly $2{\mathcal  K'}(F\tilde{F}/\Lambda^4) 
 \rightarrow 
 \theta +  {\mathcal O}(1/r^2)$, 
we obtain
 \begin{equation} \label{WittenQ}
 q_e \, = \, \theta \, q_m \,.
 \end{equation}
This reproduces the Witten effect. 
That is, under the above assumptions, we have observed that on the background with non-zero $F\tilde{F}$, the monopole becomes a dyon, which can be parameterized as, 
\begin{equation}
A^{\text{dyon}}_\mu=  \frac{1}{r}
\begin{pmatrix}
        q_e\\
    -q_m \frac{1 - \cos(\vartheta)}{\sin(\vartheta)} \sin\phi\\
    q_m \frac{1 - \cos(\vartheta)}{\sin(\theta)} \cos\phi\\
    0
    \end{pmatrix} \,
\end{equation}
with $q_e$ given by equation~\eqref{WittenQ}.

\setlength{\bibsep}{4pt}
\bibliography{references}

\end{document}